\documentclass
[
3p,times,final,authoryear]{elsarticle}

\usepackage{amsthm}

\usepackage{natbib}

\usepackage{dsfont}
\usepackage{graphicx}
\usepackage{mathrsfs}
\usepackage{amssymb}
\usepackage{amsbsy}
\usepackage{amsmath}
\usepackage{amsfonts}
\usepackage{latexsym}
\usepackage[latin1]{inputenc}
\usepackage{txfonts}
\usepackage{color}

\usepackage{hyperref}

\newcommand{\vir}[1]{``#1''}

\newcommand{\abs}[1]{\left\vert#1\right\vert}
\newcommand{\pg}[1]{\left\{#1\right\}}
\newcommand{\pq}[1]{\left[#1\right]}
\newcommand{\pt}[1]{\left(#1\right)}
\newcommand{\w}{\mathbf{w}}

\newcommand{\F}{\mathscr{F}}

\newtheorem{theorem}{\bf Theorem}

\newtheorem{proposition}{\bf Proposition}
\newtheorem{definition}{\bf Definition}

\newtheorem{Ex}{\bf Example}

\journal{}

\makeatletter
\renewcommand\@biblabel[1]{}
\renewenvironment{thebibliography}[1]
     {\section*{\refname}%
      \list{}%
           {\leftmargin0pt
            \@openbib@code
            \usecounter{enumiv}
            }%
      \sloppy
      \clubpenalty4000
      \@clubpenalty \clubpenalty
      \widowpenalty4000%
      \sfcode`\.\@m}
     {\def\@noitemerr
       {\@latex@warning{Empty `thebibliography' environment}}%
      \endlist}
\makeatother

\begin{document}

\begin{frontmatter}



\title{Bayesian adaptation}
\author{Catia Scricciolo\corref{cor1}}


\cortext[cor1]{Corresponding author.
} \ead{catia.scricciolo@unibocconi.it}




\address{Department of Decision Sciences, Bocconi University, Via Röntgen 1, 20136 Milano, Italy}

\begin{abstract}

In the need for low assumption inferential methods in infinite-dimensional settings,
Bayesian adaptive estimation via a prior distribution that does not depend on
the regularity of the function to be estimated nor on the sample size is valuable.
We elucidate relationships among the main approaches followed to design priors for
minimax-optimal rate-adaptive estimation meanwhile shedding light on the underlying ideas.
\end{abstract}


\end{frontmatter}



\section{Introduction}\label{sec:1}
Nonparametric curve estimation is a fundamental problem that
has been intensively studied in a Bayesian framework only in the last decade, with more than a ten-years delay
over the ponderous progress made in the frequentist literature where rates for point estimators
have been developed in many aspects: adaptation, sharp minimax adaptive constants etc., see, e.g., \citet{GL2012}
for recent progress in the area. Bayesian adaptive estimation is a main theme: it
accounts for designing a prior probability measure
on a function space
so that the posterior distribution
contracts at \vir{the truth} at
optimal rate, in the minimax sense, relative to the distance defining the risk.
The rate then has the desirable property of automatically adapting to the unknown
regularity level of the estimandum: the correct rate stems,
whichever the true value of the regularity parameter,
even if knowledge of it is not 
available to be exploited in the definition of the prior.
As the amount of data grows, the posterior distribution learns from the data
so that the derived estimation procedure, despite lack of knowledge of the smoothness,
performs as well as if the regularity level were known and this information could be incorporated into the prior.
In this sense, adaptation may be regarded as an oracle property of the prior distribution providing a
frequentist large-sample validation of it
and, above all, a success of Bayesian nonparametric methods for
low assumption inference in infinite-dimensional settings.

Early influential contributions to Bayesian adaptation are due to \citet{BG2003} and \citet{H2004}.
The former article deals with the prototypical problem of
adaptive estimation of the mean of
an infinite-dimensional normal distribution which is assumed to live in
a Sobolev space of unknown smoothness level; the latter provides general sufficient conditions for adaptive
density and regression estimation which are then applied to illustrate
full exact minimax-optimal rate adaptation in density and regression estimation
over Sobolev spaces
using log-spline models and full minimax-optimal rate adaptation in density estimation over Besov spaces with the
Haar basis but at the price of an extra logarithmic term.
A third breakthrough contribution is given in the article of
\citet{vdVvZ2009}, where adaptation is considered in the statistical settings of
density estimation, regression and classification by introducing as a prior for
the functional parameter a 
re-scaling of the sample paths of a smooth
Gaussian random field on $[0,\,1]^d$, $d\geq1$, by an independent gamma random variable.
These three articles are paradigmatic of the main approaches followed for Bayesian adaptation:
\begin{itemize}
\item[(a)] the approach that considers the regularity level as a hyper-parameter and puts a prior on it;
\item[(b)] the approach that puts a prior on a discrete random variable
which may represent the model dimension, the dimension of the space where the function is projected or
the number of basis functions used in the approximation;
\item[(c)] the approach based on the re-scaling of a smooth Gaussian random field.
\end{itemize}

Approach (a), which considers hierarchical models with regularity hyper-parameter,
is proposed in \citet{BG2003}, where the unknown regularity
level is 
endowed with a prior supported on at most countably many values. The overall prior is then a mixture of priors on
different models indexed by the regularity parameter 
and leads to exact optimal posterior contraction rates
simultaneously for all regularity levels.
The same philosophy is followed in \citet{S2006},
where full exact optimal rate adaptive estimation of log-densities in Sobolev ellipsoids is achieved by
considering only a finite number of competing models. In both articles,
the key 
ideas are the following:
\begin{itemize}
\item[(i)] the posterior probability of selecting a coarser model than the best one asymptotically vanishes;
\item[(ii)] the posterior distribution resulting from the prior restricted to bigger models
asymptotically accumulates on a fixed
ellipsoid in the correct space;
\item[(iii)] the posterior distribution corresponding to the restricted prior concentrates on Hellinger/$\ell^2$-balls around
the truth at optimal rate.
\end{itemize}

In both articles, full minimax-optimal rate adaptation is achieved when the prior on the regularity level can only take countably many values, while
continuous spectrum adaptation is obtained at the price of a genuine power of $n$ in \citet{BG2003}
and of an extra logarithmic factor in \citet{Lian2014}.
In the latter article, adaptation to the regularity level of the Besov space where
the true signal of a Gaussian white noise model is assumed to live is achieved, up to a log-factor,
over the full scale of possible regularity values by considering a spike-and-slab type prior, with a point mass at zero
mixed with a Gaussian distribution, on the single wavelet coefficients of the signal
and a prior on a parameter related to the regularity of the space, but the overall prior is restricted
to a fixed 
Besov ellipsoid. Another extension of \citet{BG2003} to continuous spectrum is \citet{KSvdVvZ2012}.
Also the Bayesian adaptation scheme proposed by \citet{GhosalLembervdV2003} and \citet{LembervdV07} can be ascribed to approach (a).
It puts a prior on every model of a collection, each one expressing a qualitative prior
guess on the true density, possibly a regularity parameter, and next combines
these priors into an overall prior by equipping the abstract model indices with special sample-size-dependent prior weights
giving more relevance to \vir{smaller} models, that is, those with faster convergence rates.
Illustrations include finite discrete priors based on nets and priors on finite-dimensional models
for adaptive estimation over scales of Banach spaces like Hölder spaces. A closely related
problem is that of model selection which is dealt with using similar ideas in \citet{GhosalLembervdV2008},
where it is shown that the posterior distribution gives negligible weights to models that are bigger
than the one that best approximates the true density from a given list, thus automatically selecting the optimal one.

Approach (b) that considers hierarchical models with dimension reduction hyper-parameter
is followed in \citet{H2004} and relies on the construction of a fairly simple compound prior
called \vir{sieve prior} by \citet{SW2001}. A sieve prior is a mixture of priors,
$$\Pi=\sum_{k=1}^\infty\rho(k) \Pi_k,$$
with $\rho(k)\geq0$, $\sum_{k=1}^\infty\rho(k)=1$ and, where every single prior $\Pi_k$ is supported on a space of densities $\mathscr{F}_k$ which
is typically finite-dimensional and can be represented as $\{f_{\theta}:\,\theta\in\Theta_k\}$.
As previously mentioned, the index $k$ may represent the dimension of the space where the function is projected,
the number of basis functions for the approximation or the model dimension. A sieve prior
can be thought of as generated in two steps: first the index $k$ of a model is selected with probability
$\rho(k)$, next a probability measure is generated from the chosen model $\mathscr{F}_k$ according to a prior
$\Pi_k$ on it. Such finite-dimensional models may
arise from the approximation of a collection of target densities through a set of basis functions
(e.g., trigonometric functions, splines or wavelets), where a model of dimension $k$ is
generated by a selection of $k$ basis functions. This adaptive scheme is based on
a set of assumptions such that they give control in terms of covering numbers of the
local structure of each $\Theta_k$, they guarantee the existence of a model $\mathscr{F}_{k_n}$
receiving enough prior weight $\rho(k_n)$, the existence of a density $f_{\beta_{k_n}}\in\mathscr{F}_{k_n}$ 
close to $f_0$ and of neighborhoods of this approximating density being charged enough prior
mass by 
$\Pi_{k_n}$. Several examples treated in \citet{H2004} using scales of finite-dimensional models
are covered with different priors in \citet{LembervdV07}.
Further references on adaptive curve estimation via sieve priors are \citet{S2008} and
\citet{AGR2013}. Bayesian adaptive procedures via sieve priors on the unit interval include piecewise constant and
polygonally smoothed priors based on the Dirichlet process as in \citet{S2007},
Bernstein-Dirichlet polynomials as in \citet{KvdV08}, mixtures of beta densities as in
\citet{R2010}. Other contributions clearly belonging to this category, while not being Dirichlet mixtures,
are \citet{dJvZ2010, dJvZ2012}, \citet{Ray2013} and \citet{BS2013}.
The underlying idea is that of considering a sequence of positive projection kernels so that,
at each \vir{resolution} level, the Dirichlet process filtered through the kernel results in a density.
Considering instead a \vir{convolution-type} kernel, with usual conversion from bin-width to bandwidth,
fully rate-adaptive density estimation over locally Hölder classes on the real line can be performed using finite
Dirichlet location mixtures of analytic exponential power densities as proposed by \citet{KRvdV10}.
Mixture models with priors on the mixing distribution
admitting an infinite discrete representation, like the Dirichlet process or more general
stick-breaking priors, avoid choosing a truncation level for
the number of mixing components, while updating it in a fully Bayes way is computationally intensive.
Fully rate-adaptive density estimation over Sobolev
or analytic regularity scales can be performed using Dirichlet process mixtures of Gaussian densities
as shown in \citet{S2014}. The extension to the multivariate setting is due to \citet{STG2013}.

Theoretical properties of approach (c) based on re-scaling are investigated in \citet{vdVvZ2009}, \citet{SvdVvZ2013} and \citet{CKP2014}.
Computational aspects are studied in \citet{ABPS2013}. The method is applied in
many practical articles, cf. \citet{vdVvZ2007} for some references.

Almost all the above described schemes for Bayesian adaptation yield rates with
extra logarithmic terms. The issue of whether in Bayesian nonparametrics logarithmic terms
could be removed in posterior contraction rates has been settled in the
affirmative by \citet{Gao&Zhou2014} using a novel block prior and getting a rate-optimal posterior contraction
result over a continuum of regularity levels for curve estimation over Sobolev or Besov ellipsoids
in a general framework covering various statistical settings such as density estimation,
white noise model, Gaussian sequence model, Gaussian regression and spectral density estimation.


Except for the article of \citet{H2004} and those dealing with re-scaling, all previously mentioned contributions
fall within the same approach for deriving posterior contraction rates as developed by
\citet{GGvdvV2000}, \citet{SW2001}. We expose the main underlying ideas in the case
of independent and identically distributed (i.i.d.) observations,
the case of dependent, non-identically distributed
observations adding only technical difficulties, see \citet{GvdV072} for the non-i.i.d. case.
Let $X^{(n)}$ denote the observation at the $n$th stage which consists of $n$ i.i.d. replicates
from a probability measure $P_{0}$ that possesses density
${f_0}$ with respect to (w.r.t.) some dominating measure $\lambda$
on a sample space $\mathscr{X}$. Let $\mathscr{F}:=\{P:\,P\ll\lambda\}$ be 
the collection of all probability measures on $\mathscr{X}$ that possess densities
w.r.t. $\lambda$, 
equipped with a semi-metric $d$, typically the Hellinger or
the $L^1$-distance.
\citet{GN11} 
have provided sufficient conditions
for assessing posterior contraction rates in the full scale of
$L^p$-metrics, $1\leq p\leq \infty$, in an abstract setting
using a different strategy of attack
to the problem.
Also the recent work of \citet{HRS-H2013} deals with $L^p$-metrics and
gives \vir{adapted} conditions for posterior contraction rates with the help
of modulus of continuity. The contribution of
\citet{CAS2014} is focussed on sup-norm posterior contraction rates based on yet
another approach oriented to specific statistical settings like the Gaussian white noise
model for non-conjugate priors and density estimation using priors
on log-densities or random dyadic histograms. Let $\Pi$ be a prior probability measure on $\mathscr{F}$.
The posterior probability of any Borel set $B$ writes as
\[\Pi(B\mid X^{(n)})=\frac{\int_B\prod_{i=1}^n(f_P/{f_0})(X_i)\,\Pi(\mathrm{d}P)}{\int_{\mathscr{F}} \prod_{i=1}^n(f_P/{f_0})(X_i)\,\Pi(\mathrm{d}P)},\]
where $f_P=\mathrm{d}P/\mathrm{d}\lambda$. 
A sequence $\epsilon_n\downarrow0$ such that $n\epsilon_n^2\rightarrow\infty$ is said to be (an upper
bound on) the posterior contraction rate, relative to 
$d$, if for a sufficiently large constant $M>0$ (or a slowly varying sequence $M_n\rightarrow\infty$),
\begin{equation}\label{eq3}\Pi(P:\,d(f_P,\,f_0)>
M\epsilon_n\mid X^{(n)})\rightarrow0
\end{equation}$P_0^\infty$-almost surely or in $P_0^n$-probability,
where $P_0^n$ stands for the joint law of the first $n$
coordinate projections of the infinite product probability measure $P_0^\infty$.
The intuition behind the notion of rate of convergence, as stated in
(\ref{eq3}), is that the radius $M\epsilon_n$ of a $d$-ball
around $f_0$ is large enough to prevent escape of mass as the
posterior shrinks to point mass at $P_0$.
In order to show convergence in \eqref{eq3}, it is enough
\begin{description}
\item (i) to bound above the numerator of the ratio defining the
probability in \eqref{eq3} by a term of the order $\exp(-c_1n\bar{\epsilon}_n^2)$,
\item (ii) to bound below the denominator of the ratio defining the
probability in \eqref{eq3} by a term of the order $\exp(-c_2n\tilde\epsilon_n^2)$,
\end{description}
where $c_1,\,c_2>0$ are finite suitable constants and $\bar{\epsilon}_n,\,\tilde\epsilon_n$ are sequences such that
$(\bar{\epsilon}_n\vee\tilde\epsilon_n)\rightarrow0$ and
$n(\bar{\epsilon}_n^2\wedge\tilde\epsilon_n^2)\rightarrow\infty$,
(for real numbers $a$ and $b$, we denote by $a\vee b$ their maximum and by $a\wedge b$ their minimum.
Also we write \vir{$\lesssim$} and \vir{$\gtrsim$} for inequalities valid up to a constant multiple which is universal
or inessential for our purposes).
The posterior contraction rate is then defined as
$\epsilon_n:=(\bar{\epsilon}_n\vee\tilde\epsilon_n)$.
This double sequence version of the theorem is introduced in
\citet{GvdV01}.
The exponential upper bound in (i) can be shown by considering an appropriate sieve set $\mathscr{F}_n$
which is almost the support of the prior $\Pi$, in the sense that the complement receives
exponentially small prior mass $$\Pi(\mathscr{F}_n^c)\lesssim e^{-(c_3+2)n\tilde\epsilon_n^2},$$ as proposed by \citet{Barron1988},
meanwhile controlling the complexity of $\mathscr{F}_n$ by the covering or packing number when appropriate tests exist, that is,
$$\log D(\bar\epsilon_n,\,\mathscr{F}_n,\,d)\lesssim n\bar\epsilon_n^2,$$ where
$D(\bar\epsilon_n,\,\mathscr{F}_n,\,d)$ denotes the $\bar\epsilon_n$-packing number of $\mathscr{F}_n$,
namely, the maximum number of points in $\mathscr{F}_n$ such that the distance between each pair is at least $\bar\epsilon_n$.
The exponential lower bound in (ii) is implied by the condition that Kullback-Leibler type neighborhoods of $P_{0}$
receive enough prior mass
$$\Pi(B_{\textrm{KL}}(P_{0};\,\tilde\epsilon_n^2))\gtrsim \exp{(-c_3n\tilde\epsilon_n^2)},$$
where $B_{\textrm{KL}}(P_{0};\,\tilde\epsilon_n^2):=\{P:\,
\max\{\textrm{KL}(P_{0};\,P),\,\textrm{V}_2(P_{0};\,P)\}\leq\tilde{\epsilon}_n^2\}$, for
$\textrm{KL}(P_{0};\,P):=\int {f_0}\log({f_0}/f_P)\,\mathrm{d}\lambda$ the Kullback-Leibler divergence and
$\textrm{V}_2(P_{0};\,P):=\int {f_0} |\log({f_0}/f_P)|^2\,\mathrm{d}\lambda$
the second moment of $\log({f_0}/f_P)$.
A condition which is originated from \citet{Schwartz1965}.

The analysis of the asymptotic behavior of posterior distributions in terms of
contraction rates details more comprehensively the impact of the prior on the
posterior than the analysis of the speed at which the expected squared
error between $f_0$ and the predictive density
$$\hat{f}_n(\cdot)=\int_{\mathscr F}
f_P(\cdot)\,\Pi(\mathrm{d}P\mid X^{(n)}),$$
as measured by 
the risk $\mathbb{E}^n_0[d^2(\hat{f}_n,\,f_0)]$, where $\mathbb{E}^n_0[\cdot]$ denotes expectation under
$P_0^n$, converges to zero as $n\rightarrow\infty$.
If $\epsilon_n$ is (an upper bound on) the posterior contraction rate and
the posterior probability in (\ref{eq3}) converges
to zero at least at the order $\epsilon_n^2$,
then $\epsilon_n$ is (an upper bound on) the rate of convergence of the
Bayes' estimator, 
provided $d$ is bounded and its square convex.
The posterior contraction rate is related to the minimax rate of convergence over the density function class which
$f_0$ belongs to. Let $\mathscr F_\beta$ denote a density function class
indexed by a parameter $\beta$ related to the regularity of its elements.
\begin{definition}
A positive sequence 
$\epsilon_{n,\,\beta}\rightarrow0$ 
is said to be the minimax rate of convergence over
$(\mathscr F_\beta,\,d)$ if there exist universal constants $0<c,\,C<\infty$,
possibly depending on the regularity parameter $\beta$,
such that the minimax risk over $\mathscr F_\beta$, that is,
$\inf_{\hat f_n}\sup_{f\in \mathscr F_\beta}\mathbb{E}_f^n[d^2(\hat{f}_n,\,f)]$, where
$\hat f_n$ stands here for any density estimator based on $n$ observations, satisfies
\[c \leq\liminf_{n\rightarrow\infty}\epsilon_{n,\,\beta}^{-2}\, \inf_{\hat f_n}\sup_{f\in \mathscr F_\beta}\mathbb{E}_f^n[d^2(\hat{f}_n,\,f)]
\leq\limsup_{n\rightarrow\infty}\epsilon_{n,\,\beta}^{-2}\, \inf_{\hat f_n}\sup_{f\in \mathscr F_\beta}\mathbb{E}_f^n[d^2(\hat{f}_n,\,f)]\leq C,
\]
where 
$\mathbb{E}^n_f[\cdot]$ denotes expectation under $P^n_f$.
An estimator $\hat f_n^*$ is said to be adaptive
in the minimax sense over the collection of function spaces
$\{\mathscr{F}_\beta,\,\beta\in B\}$ if there exists a constant $0<C_1<\infty$, possibly depending on $\beta$, such that
\[\forall\,\beta>0,\quad\sup_{f\in \mathscr F_\beta}\mathbb{E}_f^n[d^2(\hat{f}_n^*,\,f)]\leq C_1\epsilon_{n,\,\beta}^2.\]
\end{definition}
Since the rate of convergence of an estimator cannot be faster than the minimax rate over the considered density function
class, the posterior contraction rate cannot be faster than the minimax rate.
So, if the posterior distribution achieves the minimax rate, then also the Bayes' estimator
$\hat{f}_n$ has minimax-optimal convergence rate and is adaptive.
Furthermore, by taking the center of the smallest ball accumulating at least $3/4$ of the posterior mass gives a
point estimator with the same rate of convergence as the posterior contraction
rate without requiring convexity of $d^2$, see Section~4 in \citet{BG2003}.
The study of posterior contraction rates may thus
play an ancillary role in allowing to appeal to 
general theoretical results, see Theorem 2.1 and Theorem 2.2 in \citet{GGvdvV2000}
or Theorem 2 and Theorem 4 in \citet{SW2001}.

In this overview, while trying to convey the main underlying ideas,
we attempt at providing an account of the state of the art on Bayesian adaptation and an update of existing
monographs on the theme like the one by \citep[Ch. 2]{G2010} and
the dissertation by \citet{S2013} of which we point out the contributions of
Ch.~3 devoted to curve estimation using random series priors. For a variety of reasons, here we focus on Bayesian adaptation by mixtures,
this having the two-fold meaning of modeling the data-generating density
by mixtures and of using compound priors that are themselves
mixtures of priors like sieve priors. We try to set up a unifying framework useful for
understanding the large-sample behavior of commonly used priors
as well as possibly being the starting point for the development of new results. Interest in mixtures
is doubly motivated by the fact that they naturally arise in many contexts as models for
observations of phenomena with multiple underlying factors and
by their flexibility, due to which they may provide good approximation schemes for function estimation.
For instance, the combination of a Gaussian convolution kernel with a Dirichlet process prior
constitutes one of the most popular Bayesian schemes for density estimation on the real line.
As pointed out in \citet{S2013}, results concerning the approximation of densities by
Gaussian mixtures pave the way to the achievement of results on the estimation of density
derivatives which are important because involved in relevant statistical quantities such as the score function and the Fisher information. Another important problem for which mixtures are well-suited
is that of estimating multivariate (possibly anisotropic) densities, see \citet{STG2013}.
A closely related problem is that of the estimation of mixing distributions. While the problem has been
extensively studied from a frequentist perspective using deconvolution kernel-type estimators, Bayesian nonparametric
deconvolution has been hardly investigated so far, except for the
recent article of \citet{Ng2013} and the manuscripts by \citet{SPMC2013}, who derive
adaptive posterior convergence rates for Bayesian density deconvolution with supersmooth errors,
and by \citet{DRRS2014} where both the ordinary and the supersmooth cases are treated
in a fully Bayes as well as in an empirical Bayes context.


In Section~\ref{sec:2}, we provide a survey of results on Bayesian adaptation for
the most popular schemes for density estimation by mixtures.
For a more comprehensive overview of the diverse contexts and fields of application
of mixture models, the reader may consult \citet{MMR2005}.
The focus of the article is on fully Bayes adaptation techniques, but
some lights on empirical Bayes adaptation methods and on adaptive nonparametric credibility regions
are shed in Section~\ref{sec:3}.


\section{Bayesian adaptation by mixtures}\label{sec:2}
Mixtures of probability distributions naturally arise in many contexts as models for
observations of phenomena with multiple latent factors, so that
modeling by mixtures is well motivated in such situations.
On other side, in a Bayesian set-up, mixtures can be the
building block for constructing priors on spaces of densities
using a model-based approach since, by endowing the mixing
distribution of a mixed density with a probability measure, a prior distribution can be induced
on a space of probability measures possessing densities w.r.t. some dominating measure.
Furthermore, a well-chosen mixture model may
provide an approximation scheme for density estimation
resulting in minimax-optimal convergence rates. This approach, which has the virtue of combining
conceptual simplicity of the scheme with flexibility of the model due to the wide range of
possible choices for the kernel, has been initiated by
\citet{F}, who used a Dirichlet process prior
for the mixing distribution and derived the
expressions for the resulting posterior distribution
and Bayes' density estimator or predictive density, see also \citet{Lo1984}.

Given a kernel $K$, namely, a jointly measurable mapping from
$\mathscr{X}\times\Theta$ to $\mathbb{R}^+$ such that,
for every fixed $\theta\in\Theta$, $K(\cdot;\,\theta)$ is a probability density on $\mathscr{X}$ w.r.t.
$\lambda$,
a way for defining a prior is that of modeling the random probability density w.r.t.
$\lambda$ as
\begin{equation}\label{kernel}
x\mapsto f_P(x)=\int_\Theta K(x;\,\theta)\,P(\mathrm{d}\theta),
\end{equation}
where the mixing probability measure $P$ is endowed with a prior $\Pi$.
So, conditionally on $f_P$, the observations $X_i$ are i.i.d. according to
$f_P$. A way to structurally describe
observations from a kernel mixture prior is via the following hierarchical model:
\begin{equation*}\label{DPMGaussian}
\begin{split}
X_i\mid \theta_i,\,P\overset{\textrm{ind}}\sim &\, K(\cdot;\,\theta_i),\quad i=1,\,\ldots,\,n,\\
\theta_i\mid P\overset{\textrm{iid}}\sim&\, P,\qquad\quad\,\, i=1,\,\ldots,\,n,\\
P\sim&\,\Pi.
\end{split}
\end{equation*}
In the original formulation of \citet{F}, 
the combination of a Gaussian kernel and a Dirichlet process has been proposed
for density estimation on the real line
and the mixture model is called Dirichlet process mixture of Gaussian densities.
This is the most popular Bayesian scheme for density estimation on the real line,
but the need may arise for the use of different kernels because
the empirical distribution of many phenomena fails to conform to a Gaussian distribution,
thus leading to the search
for other models. An alternative when the discrepancy lies in the tails can be represented
by exponential power distributions, where the tail thickness is governed by a shape parameter.
For example, the normal-Laplace distribution, resulting from the convolution of independent normal and Laplace components,
behaves like the normal in the middle of its range and like the Laplace in the
tails. 
Its use in the study of high frequency price data is pointed out in \citet{Reed}.
Rates of contraction for density estimation using Dirichlet mixtures of exponential power densities are derived in
\citet{S2011}. Another possibility is that of employing a kernel belonging to the family of (symmetric) stable laws,
which includes the Cauchy. 
Unlike exponential power distributions, these distributions have heavy
(polynomially decaying) tails and arise in many applications.
For compactly supported data, the combinations of a Dirichlet distribution with Bernstein polynomials \citep{Petrone99},
triangular densities \citep{PerMeng2001, McVinishRousseauMengersen2009}, histograms or polygons \citep{S2007}
have been suggested. Some of them are illustrated in the examples below.

In this survey, we are mostly interested in nonparametric mixtures, that is, in the case where the number 
of the underlying components is unknown and 
infinite, and want to consider their theoretical properties.
Given a random sample of i.i.d. observations $X_1,\,\ldots,\,X_n$ from the \vir{true} distribution $P_{0}$, we are interested in studying
frequentist asymptotic properties of the posterior distribution as the sample size tends to infinity, the 
focus being on adaptation to unknown smoothness. 
Consider observations $X_1,\,\ldots,\,X_n$ from a density $f_0$ on $\mathbb{R}$, 
or on some subset thereof, belonging to a model $\mathscr{F}_\beta$. 
For example, $\mathscr{F}_\beta$ could be the space of density functions on $[0,\,1]$ that are Hölder $\beta$-smooth.
Recall that, for $\beta>0$, a density (or, more generally, a function) $f$ defined on a set $\mathscr{X}\subseteq\mathds{R}$ is said to be Hölder
$\beta$-\emph{smooth} if it is differentiable up to the order
$\lfloor\beta\rfloor:=\max\{i\in\mathbb{N}:\,i<\beta\}$ 
and the derivative $f^{(\lfloor\beta\rfloor)}$ is (uniformly) Hölder continuous with exponent
$\beta-\lfloor\beta\rfloor$,
\begin{equation}\label{eq:Holder}
|{f^{(\lfloor\beta\rfloor)}(x)-f^{(\lfloor\beta\rfloor)}(y)}|\leq
L\abs{x-y}^{\beta-\lfloor\beta\rfloor},\quad\forall\,x,\,y\in\mathscr{X},
\end{equation}
where $L>0$ is a finite constant, possibly depending on $f$ and $\beta$.
For later use, we introduce the notation
$$[f]_\beta:=\sup_{x\neq y}\frac{|{f^{(\lfloor\beta\rfloor)}(x)-f^{(\lfloor\beta\rfloor)}(y)}|}{\abs{x-y}^{\beta-\lfloor\beta\rfloor}}$$
to denote the smallest constant for which \eqref{eq:Holder} is satisfied.
Let $C^\beta([0,\,1])$ stand for the class
of (Lebesgue) densities on $[0,\,1]$ that are Hölder $\beta$-smooth.
Consider a scale of models $\{\mathscr{F}_\beta,\,\beta\in B\}$.
The value of the regularity parameter $\beta$ of $f_0$ is typically unknown. The problem is that of designing
a prior supported on $\bigcup_{\beta\in B}\mathscr{F}_\beta$ such that the
posterior, hence the entailed Bayes' estimator, has the remarkable fine property of
being \emph{self-adaptive} to $\beta$, in the sense that, as the value of $\beta$
varies, one need not change the prior to guarantee that the
corresponding posterior achieves the minimax-optimal contraction rate simultaneously over all 
classes of the collection. The rate of convergence thus has the
property of adapting automatically to the unknown smoothness level
$\beta$ of $f_0$. In other terms, the correct rate stems, whatever
the true value of $\beta$, even if $\beta$ is not involved in the
definition of the prior.  
Henceforth, $\epsilon_{n,\,\beta}$ stands for the minimax-optimal rate of convergence relative to
the $L^1$-metric over the class $\mathscr F_\beta$. 
\begin{definition}
The posterior distribution corresponding to a prior
measure $\Pi$ on $\bigcup_{\beta\in B}\mathscr{F}_{\beta}$
concentrates adaptively over 
$\{\mathscr{F}_{\beta},\, \beta\in B\}$ 
if, for some finite constant $M>0$ (or some slowly varying sequence $M_n\rightarrow\infty$),
$$\forall\,\beta\in B,\quad\sup_{f_0\in\mathscr{F}_{\beta}}
\mathbb{E}_0^n[\Pi(P:\,\|f_P-f_0\|_1> M\epsilon_{n,\,\beta}\mid X^{(n)})]\rightarrow0.$$ 
\end{definition}

As mentioned in Section~\ref{sec:1}, approach (b), which is based on hierarchical models with dimension reduction hyper-parameter, 
relies on the construction of the so-called sieve priors.
A sieve prior is a mixture of priors $\sum_{k=1}^\infty\rho(k)\Pi_k$,
where $\Pi_k$ is
supported on some set of densities $\mathscr{F}_k$ with generic element
$f_k$ that can be a kernel mixture. The overall prior
$\Pi=\sum_{k=1}^\infty\rho(k)\Pi_k$ induces a prior on $\F$ which (almost surely) selects probability measures with densities $f_k$.
The choice of the densities $f_k$ is motivated by the fact that they possess some approximation property
for \vir{regular} densities, 
relative to some $L^p$-metric, $1\leq p\leq\infty$. In fact,
if $\rho(k)$ is positive for all but finitely many $k$ and $\Pi_k$
is fully supported on the $(k-1)$-dimensional standard simplex
$\Delta^{k-1}:=\{\w_k\in \mathbb{R}^k:\,\sum_{j=1}^kw_{j,\,k}=1,\,w_{j,\,k}\geq0 \mbox{ for }j=1,\,\ldots,\,k\}$,
then every probability measure with density $f_P$ which is the
$L^p$-limit of a sequence of densities $f_k$, that is,
$\lim_{k\rightarrow\infty}\|f_k-f_P\|_p=0$, is
in the $L^p$-support of $\Pi$. 
The approximation property of densities $f_k$ is crucial to assess the prior concentration rate
$\tilde{\epsilon}_n$, which is a main determinant of the posterior contraction rate
at \vir{regular} densities. In fact, the main challenge when proving adaptation
lies in finding a finite mixing distribution, with a suitable number of support points, such that the corresponding
kernel mixture approximates the sampling density, in the Kullback-Leibler divergence,
with an error of the correct order.
Mixtures are constructed so that their approximation properties guarantee that,
under natural conditions on the priors
of the hierarchy, the prior mass in
Kullback-Leibler type neighborhoods around the sampling density $f_0$ is bounded below by the
probability of the mixing weights taking values in a simplex of appropriate dimension,
say $(k_0-1)$, depending on the true value of the regularity parameter $\beta$ and the approximation error $\epsilon$,
\[\Pi(B_{\textrm{KL}}(P_0;\,\epsilon^2))\geq \rho(k_0)\Pi_{k_0}(N(f^*_{k_0};\,\epsilon^2)),\]
where $N(f^*_{k_0};\,\epsilon^2)\subseteq \Delta^{k_0-1}$ is an 
Euclidean ball centered at the best approximation $f^*_{k_0}$ to $f_0$ in $\mathscr{F}_{k_0}$.
This crucial step can be better understood from the following examples.

\begin{Ex}\emph{(}Random histograms and Bernstein polynomials\emph{)}.
\emph{Random histograms are a common nonparametric model-based mixture prior.
For every $k\in\mathbb{N}$, let
$\{A_{1,\,k},\,\ldots,\,A_{k,\,k}\}$ be the partition of 
$[0,\,1]$ into $k$ intervals (\emph{bins}) of equal length
$\lambda(A_{j,\,k})=1/k$ for $j=1,\,\dots,\,k$,
where $A_{1,\,k}=[0,\,1/k]$ and $A_{j,\,k}=((j-1)/k,\,j/k]$ for $j=2,\,\ldots,\,k$. Given the number $k$ of bins,
for any $\w_k=(w_{1,\,k},\,\ldots,\,w_{k,\,k})\in\Delta^{k-1}$,
let the $k$-regular histogram be defined as
$h_{\w_k}(x)=\sum_{j=1}^k w_{j,\,k}k
\mathds{1}_{A_{j,\,k}}(x)$, $x\in[0,\,1]$, where the $w_{j,\,k}$
are the mixing weights for the densities $k
\mathds{1}_{A_{j,\,k}}$, with $\mathds{1}_{A_{j,\,k}}$ the
indicator function of the cell $A_{j,\,k}$.
The prior can be constructed by randomizing the number $k$ of bins and the mixing weights 
$\w_k$. First the index $k$ is selected with probability $\rho(k)$, next a probability measure
is generated from the chosen model $h_{\w_k}$ according to a prior $\Pi_k$ for $\w_k$,
the prior $\Pi_k$ being typically chosen to be a
Dirichlet distribution on the $(k-1)$-dimensional simplex $\Delta^{k-1}$ with parameters
$\alpha_{1,\,k},\,\ldots,\,\alpha_{k,\,k}$, i.e., $\Pi_k=
\text{Dir}_k(\alpha_{1,\,k},\,\ldots,\,\alpha_{k,\,k})$. The mixing weights
may be thought of the form $w_{j,\,k}=
P(A_{j,\,k})$, where $P$ is a random probability measure
distributed according to a Dirichlet process with base measure
$\alpha$, in symbols, $P\sim\textrm{DP}(\alpha)$.
A \emph{piecewise constant prior} can be structurally described as follows.
Defined the function
$N_k(\cdot):=\sum_{j=1}^kk\mathds{1}_{A_{j,\,k}}(\cdot)$, the hierarchical model is
\begin{equation}\label{DPMGaussian}
\begin{split}
X_i\mid k,\,P,\,\theta_i\,\overset{\textrm{ind}}\sim & \, N_k(\theta_i)=k\mathds{1}_{A_{j(\theta_i),\,k}}(\theta_i), \quad i=1,\,\ldots,\,k,\\
\theta_i\mid k,\,P\overset{\textrm{iid}}\sim & \, P,\hspace*{2,99cm} i=1,\,\ldots,\,k,\\
P\mid k\sim &\, \Pi_k\\
k\sim &\, \rho,
\end{split}
\end{equation}
where $j(x)$ identifies the bin containing the point
$x$, i.e., $A_{j(x),\,k}\ni x$.
We now clarify how, conditionally on the number $k$ of bins,
the random density $h_{\w_k}$ can be written as a kernel mixture in the form \eqref{kernel}.
Taken 
$P\sim\textrm{DP}(\alpha)$,
for every $k$, consider the discretization $\alpha^{(k)}:=\sum_{j=1}^k\alpha_{j,\,k}\delta_{j/k}$ of the base measure $\alpha$,
with $\alpha_{j,\,k}:=\alpha(A_{j,\,k})$ for $j=1,\,\ldots,\,k$.
The measure $\alpha^{(k)}$ defines a random probability measure $\pi_k:=\sum_{j=1}^kw_{j,\,k}\delta_{j/k}$
supported on $\{1/k,\,\ldots,\,(k-1)/k,\,1\}$, with random weights $w_{j,\,k}=P(A_{j,\,k})$ having prior expectation
$\mathbb{E}[w_{j,\,k}]=\alpha_{j,\,k}/\alpha([0,\,1])$, for $j=1,\,\ldots,\,k$.
A piecewise constant prior is then 
the probability distribution of the random density
$f_P(\cdot)=\sum_{k=1}^\infty\rho(k)h_{\w_k}(\cdot)$, where
$$h_{\w_k}(\cdot)=\int_0^1k\mathds{1}_{A_{k\theta,\,k}}(\cdot)\pi_k(\mathrm{d}\theta)$$ is a mixture as in \eqref{kernel}, with kernel
$K(\cdot;\,\theta)=k\mathds{1}_{A_{k\theta,\,k}}(\cdot)$.
The Bayes' estimator 
yielded by a piecewise constant prior has the
following structure
$$
\hat{f}_n(\cdot)=\sum_{k=1}^\infty\rho(k \mid X^{(n)})
\sum_{j=1}^k\mathbb{E}[w_{j,\,k}\mid k,\,X^{(n)}]k\mathds{1}_{A_{j,\,k}}(\cdot),$$
which evidentiates that the posterior expected density is still a histogram with updated weights,
see equation (3) in \citet{S2007} for the complete explicit expression of $\hat{f}_n$.
Consistency of the posterior distribution of a piecewise constant Dirichlet prior
concentrated on the $k_n$-regular dyadic histograms is addressed in \citet{BSW1999}, see also
\citet{Barron1988}. The main idea is to show that the prior
satisfies \citet{Schwartz1965}'s prior positivity condition.
The posterior is
consistent in the Hellinger or the $L^1$-metric at any density $f_0$ such that $\textrm{KL}(P_{0};\,\lambda)<\infty$, for $k_n=O(n/\log n)$ and all $\alpha_{j,\,k_n}=a^n(1-a)$, with
$a\in(0,\,1)$.
Bayesian adaptive density estimation via a piecewise
constant prior is studied in \citet{S2007}.
The capability of the posterior distribution to achieve minimax-optimal contraction rates, possibly
up to a logarithmic factor, depends on the
approximation error of a density by histograms:
the sup-norm $\|\cdot\|_\infty$ approximation error of a density
$f_0\in C^\beta([0,\,1])$ by a $k$-regular histogram-shaped
density is of the order $k^{-(\beta\wedge1)}$,
which is at most only proportional to the inverse of the bin-width $k^{-1}$.
For $\beta\in(0,\,1]$, we have $\|{f_0-h_{\w^0_k}}\|_\infty\leq
L_0k^{-\beta}$, where 
$h_{\w^0_k}=\sum_{j=1}^k kw_{j,\,k}^0\mathds{1}_{A_{j,\,k}}$, with $w_{j,\,k}^0=\int_{A_{j,\,k}}f_0\,\mathrm{d}\lambda$,
is the histogram-shaped density based on $f_0$. Thus, as stated in Proposition \ref{prop1} below, piecewise constant priors can achieve minimax rates, up to a logarithmic factor,
only up to Hölder regularity $1$. It is known from the vast literature on density estimation on the unit interval $[0,\,1]$ that the minimax $L^p$-risk
$R_n^{(p)}(H(\beta,\,L)):=\inf_{\hat f_n}\sup_{f\in H(\beta,\,L)}\{\mathbb{E}_f^n[\|\hat f_n-f\|_p^2]\}^{1/2}$
over Hölder smoothness classes satisfies
$$R_n^{(p)}(H(\beta,\,L))\asymp L^{1/(2\beta+1)}\times\left\{
  \begin{array}{ll}
    n^{-\beta/(2\beta+1)}, & \hbox{for $1\leq p< \infty$,} \\
    (n/\log n)^{-\beta/(2\beta+1)}, & \hbox{for $\quad\,\,\,\, p=\infty$,}
  \end{array}
\right.$$
where $H(\beta,\,L)$ denotes the Hölder
class of order $\beta$, consisting of densities $f$ on $[0,\,1]$
such that the derivative $f^{(\lfloor\beta\rfloor)}$ exists and $[f]_\beta+\|f\|_\infty\leq L$.
Note that, except for the case where $p=\infty$, the rate does not depend on $p$.
In what follows, the previously introduced sequence $\epsilon_{n,\,\beta}$ specifies as $\epsilon_{n,\,\beta}={n}^{-\beta/(2\beta+1)}$ for $L^p$-metrics, when $1\leq p<\infty$.
Before reporting a result on adaptation, it is worth mentioning some recent findings on non-adaptive posterior
contraction rates in $L^p$-metrics for random dyadic histograms with a sample size-dependent number
$k_n=2^{J_n}=(n/\log n)^{1/(2\beta+1)}$ of bins, for densities
of Hölder regularity $\beta\in(1/2,\,1]$. \citet{GN11} obtain the minimax rate
$\epsilon_{n,\,\beta}$, up to a logarithmic factor, for $L^p$-metrics, with $p\in(0,\,2]$; while \citet{CAS2014}
gets the exact minimax sup-norm rate $(n/\log n)^{-\beta/(2\beta+1)}$.
\begin{proposition}[\citet{S2007}]\label{prop1}
Let the density $f_0\in C^\beta([0,\,1])$,
with $\beta\in(0,\,1]$, be bounded away from zero on $[0,\,1]$. Let $\Pi$ be a piecewise
constant prior, with $B_1e^{-\beta_1 k}\leq\rho(k)\leq B_2e^{-\beta_2
k}$ for all $k\in\mathbb{N}$ and constants
$B_1,\,B_2,\,\beta_1,\,\beta_2>0$, and with the base measure
$\alpha$ of the Dirichlet process possessing a continuous and positive
density on $[0,\,1]$. Then, for
$M>0$ large enough, $\Pi(P:\, \|f_0-f_P\|_1> M\epsilon_{n,\,\beta}(\log n)^{\beta/(2\beta+1)}\mid
X^{(n)})\rightarrow0$ with $P_0^\infty$-probability one. 
Consequently, $\mathbb{E}^n_0[\|\hat{f}_n-f_0\|_1^2]=O(\epsilon_{n,\,\beta}^2(\log n)^{2\beta/(2\beta+1)})$.
\end{proposition}
Since piecewise constant priors can attain minimax rates in the $L^1$-metric only up to Hölder regularity $1$,
they are not appropriate for estimating smoother than Lipschitz densities.
One may compare the performance of random histograms with that of random Bernstein polynomials.
A Bernstein-Dirichlet prior has the same structure as a piecewise constant prior described in \eqref{DPMGaussian}, but
with $N_k(\cdot):=\sum_{j=1}^kj\mathds{1}_{A_{j,\,k}}(\cdot)$ and $X_i\mid k,\,P,\,\theta_i\,
\overset{\textrm{ind}}\sim
\textrm{Beta}(N_k(\theta_i),\,k-N_k(\theta_i)+1)$. The Dirichlet process mixture of Bernstein polynomials as a nonparametric prior is
introduced in \citet{Petrone99}. 
Weak and Hellinger posterior consistency are investigated in \citet{PW2002}, while convergence rates relative to the Hellinger or the
$L^1$-distance are analyzed in \citet{G2001}. Although the sub-optimal rate found by \citet{G2001} for estimating twice
continuously differentiable densities is only an upper bound
on the posterior contraction rate, it indicates, following from Proposition \ref{prop1}, that
random histograms, despite their simple structure,
possess better approximation properties than random Bernstein polynomials, whose use in
Bayesian adaptive estimation of densities with Hölder regularity $\beta\in(0,\,2]$ has been considered by \citet{KvdV08}.
They find the sub-optimal rate 
$n^{-\beta/(2\beta+2)}$,
up to a logarithmic factor. As remarked by the authors themselves,
sub-optimality of the rate can be understood from sub-optimality of Bernstein polynomials as an approximation scheme.
In fact, in terms of approximation of Hölder regular functions, they are sub-optimal in yielding an approximation error
of the order $k^{-\beta/2}$, whereas polynomials of degree $k$ of best approximation have an error of the order $k^{-1}$ only.
We incidentally note that, as discussed in the following example, the same sub-optimality phenomenon is observed
for polygons which, in principle, are introduced to overcome limitations of histograms,
but turn out to suffer from the same deficiency when $\beta>2$.
The authors employ \emph{coarsened} Bernstein polynomials to get the nearly
optimal rate $\epsilon_{n,\,\beta}(\log n)^{(4\beta+1)/(4\beta+2)}$ for
densities of Hölder regularity $\beta\in(0,\,1]$.
Adaptation in the Hellinger metric over the full scale of Hölder classes of regularity $\beta>0$
can be achieved by using suitably constructed mixtures of beta densities, see \citet{R2010}.
}
\end{Ex}

\begin{Ex}\emph{(}Random polygons\emph{)}. \emph{A \emph{polygonally smoothed
prior}, introduced in \citet{S2007}, is a model-based
hierarchical prior having the same structure as a piecewise constant prior, but
with a continuous polygon-shaped, in lieu of a histogram-shaped, conditional density of the
observations. The polygon can be regarded as the result of a histogram
smoothing performed by joining the heights at mid-bin points $c_{j,\,k}=(j-1/2)/k$, for $j=1,\,\ldots,\,k$, with
straight lines,
\[p_{\w_k}(x)=w_{1,\,k}k\mathds{1}_{A_{1,\,k}^-}(x)+\sum_{j=1}^{k-1}
[k(c_{j+1,\,k}-x)w_{j,\,k}+k(x-c_{j,\,k})w_{j+1,\,k}]k\mathds{1}_{A_{j,\,k}^+\cup
A^-_{j+1,\,k}}(x)+w_{k,\,k}k\mathds{1}_{A_{k,\,k}^+}(x),\quad
x\in[0,\,1],\]
where, for every $j=1,\,\ldots,\,k$, the symbols $A_{j,\,k}^-$ and $A_{j,\,k}^+$ stand for the left and right equal length sub-intervals of
$A_{j,\,k}$, respectively.
Any 
density $f_0\in C^\beta([0,\,1])$
can be uniformly approximated by a $k$-regular polygon-shaped
density $p_{\w_k^0}$ based on $f_0$ with an error of the order
$k^{-(\beta\wedge1)}$, that is, $\|f_0-{p_{\w_k^0}}\|_\infty=O(k^{-(\beta\wedge1)})$.
If 
$f_0$ is Hölder $\beta$-regular, with $\beta\in(1,\,2]$, the
approximation error near the endpoints of $[0,\,1]$, where $p_{\w_k^0}$
inherits the structure of a histogram, is only of the order
$k^{-1}$, as for Lipschitz densities. Thus, extra regularity
conditions on $f_0'$, aimed at compensating for the poor approximation quality of the polygon
$p_{\w_k^0}$ near the unit interval endpoints,
can be considered to
guarantee the correct order of the approximation
error. For $\beta>1$,
possible boundary conditions on $f_0'$ are
$(\mathrm{BC1})$ $f'_0(x)=a_0x^p+o(x^p)$, as
$x\downarrow0$, and $(\mathrm{BC2})$ $f'_0(x)=b_0(1-x)^q+o((1-x)^q)$ as $x\uparrow1$,
where $a_0,\,b_0\in\mathbb{R}$ and $(\beta-1)\leq p,\,q<\infty$, see also \citet{S2007}.
\begin{proposition}\label{prop3}
Let the density $f_0\in C^\beta([0,\,1])$, with $\beta>0$.
For $\beta>1$, suppose further that $f_0'$ satisfies the boundary conditions $\mathrm{(BC1)}$ and $\mathrm{(BC2)}$.
Then, $\|f_0-p_{\w_k^0}\|_\infty=O(k^{-(\beta\wedge2)})$.
\end{proposition}
This approximation result, whose proof is deferred to \ref{App},
is the key ingredient for proving that the posterior distribution corresponding
to a polygonally smoothed prior is rate-adaptive over 
a scale of H\"older classes of regularity $\beta\in(0,\,2]$. 
\begin{theorem}\label{th1}
Let the density $f_0\in C^\beta([0,\,1])$, with $\beta\in(0,\,2]$, and
$1/f_0\in L^1(\lambda)$. For $\beta\in(1,\,2]$, suppose further that
$f_0'$ satisfies the boundary conditions $\mathrm{(BC1)}$ and $\mathrm{(BC2)}$.
Let $\Pi$ be a polygonally smoothed prior, with
$B_1e^{-\beta_1 k\log k}\leq\rho(k)\leq B_2e^{-\beta_2 k\log k}$ for all
$k\in\mathbb{N}$ and constants $B_1,\,B_2,\,\beta_1,\,\beta_2>0$, and with
the base measure $\alpha$ of the Dirichlet process having
a continuous and positive density on $[0,\,1]$. Then, for a
sufficiently large constant $M>0$,
$\mathbb{E}^n_0[\Pi(P:\, \|f_P-f_0\|_1> M\epsilon_{n,\,\beta}(\log n)^{\beta/(2\beta+1)}
\mid
X^{(n)})]\rightarrow0$.  
Consequently,
$\mathbb{E}^n_0[\|\hat{f}_n-f_0\|_1^2]=O(\epsilon_{n,\,\beta}^2(\log n)^{2\beta/(2\beta+1)})$.
\end{theorem}
Estimating any 
density 
of Hölder regularity $\beta\in(0,\,2]$ with the
Bayes' estimator entailed by a polygonally smoothed prior we may pay, at
most, a price of a $(\log n)^{\beta/(2\beta+1)}$-factor,
the convergence rate 
being \emph{self-adaptive} to 
$\beta$: 
as the regularity
parameter $\beta$ varies with $f_0$, one need not change the prior to guarantee that the
Bayes' estimator achieves, up to a multiplicative
logarithmic term, the minimax rate of convergence
over 
a scale of H\"older classes of regularity $\beta\in(0,\,2]$. 
For any $\beta>2$, instead, the error made in uniformly
approximating a density $f_0\in C^\beta([0,\,1])$ by the $k$-regular polygon-shaped
density $p_{\w_k^0}$ is only of the order $k^{-2}$
and the posterior contraction rate we find is $(n/\log n)^{-2/5}$ as for densities that are
only twice differentiable.
Hereafter, we show that the minimax $L^1$-rate $n^{-3/7}$ for the H\"older smoothness class
$C^3([0,\,1])$
is a lower bound on the contraction rate of the posterior distribution of a polygonally smoothed prior
at densities in $C^3([0,\,1])\cap \mathscr F_3$, where the subclass
$$\mathscr F_3=\left\{f|\,f:\,[0,\,1]\rightarrow\mathbb{R}^+,\, \|f\|_1=1,\, f''\mbox{ bounded away from  $0$ on an interval $I\subset (0,\,1)$ and $f'''$ bounded}\right\}$$ has been employed for an analogous purpose by \citet{McVinishRousseauMengersen2005, McVinishRousseauMengersen2009}.
We consider densities in $C^3([0,\,1])\cap \mathscr F_3$
that also satisfy the above boundary conditions $(\mathrm{BC1})$ and $(\mathrm{BC2})$.
An example of such a density is $f_0(x)=x^4-4x^3/3+17/15$, $x\in[0,\,1]$, \citep[see, e.g.,][Remark 5]{S2007}.
\begin{proposition}\label{prop4}
Let the density $f_0\in C^3([0,\,1])\cap \mathscr F_3$ satisfy the boundary conditions $(\mathrm{BC1})$ and $(\mathrm{BC2})$
and $1/f_0\in L^1(\lambda)$. Let $\Pi$ be a polygonally smoothed prior, with
$B_1e^{-\beta_1 k\log k}\leq\rho(k)\leq B_2e^{-\beta_2 k\log k}$ for all
$k\in\mathbb{N}$ and constants $B_1,\,B_2,\,\beta_1,\,\beta_2>0$, and with
the base measure $\alpha$ of the Dirichlet process having
a continuous and positive density on $[0,\,1]$. Then, 
$\Pi(P:\, \|f_P-f_0\|_1\leq n^{-3/7}
\mid
X^{(n)})\rightarrow0$  in $P_0^n$-probability.
\end{proposition}
The assertion implies that the minimax $L^1$-rate $\epsilon_{n,\,3}=n^{-3/7}$
has a too small order of magnitude to be the radius of an $L^1$-ball around $f_0$ that is
able to capture almost all the mass when the posterior weakly converges to
a point mass at $P_0$. Thus, random polygons can only get minimax rates of convergence,
up to a logarithmic factor, over a scale of H\"older classes
up to regularity $\beta=2$: they are not appropriate for estimating
smoother than twice differentiable densities because they are structurally not
able to exploit additional regularity.}

\emph{It is interesting to investigate the relationship between the Bayes' estimator 
of a polygonally smoothed prior and a frequentist counterpart,
 the so-called \emph{smooth Barron-type density estimator} proposed by
\citet{BBBV2002}:
\[f_n^P(x)\coloneqq(1-a_n)p_{\w_{k_n}^{\mu_n}}(x)+a_n,\quad x\in[0,\,1],\]
where, for $k_n\in\mathbb{N}$ such that
$\lim_{n\rightarrow\infty} k_n=\infty$ and $\lim_{n\rightarrow\infty} n/k_n=\infty$,
the sequence $(a_n)_{n\geq1}$ has generic term
$a_n=(1+n/k_n)^{-1}\rightarrow0$ and $p_{\w_{k_n}^{\mu_n}}$ is the $k_n$-regular frequency
polygon constructed with weights $w_{j,\,k_n}^{\mu_n}$ that are the relative
frequencies of the observations falling into the bins $A_{j,\,k_n}$, that is,
$w_{j,\,k_n}^{\mu_n}\coloneqq\mu_n(A_{j,\,k_n})$, for $j=1,\,\ldots,\,k_n$, where $\mu_n$ stands
for the empirical measure associated with the sample $X_1,\,\ldots,\,X_n$, i.e.,
$\mu_n(A)={n}^{-1}\sum_{i=1}^n\mathds{1}_A(X_i)$ for every measurable set $A$.
Thus, $f_n^P$
is a convex combination of the frequency polygon $p_{\w_{k_n}^{\mu_n}}$ and
the uniform density on $[0,\,1]$ and, as the sample size $n$ increases,
it shrinks towards the frequency polygon which converges pointwise to $f_0$. The smooth Barron-type density estimator
is a modification of the histogram-based \emph{Barron estimator} \citep{Barron1988}
\begin{equation}\label{eq:Bestim}
f_n^B(x)\coloneqq(1-a_n)h_{\w_{k_n}^{\mu_n}}(x)+a_n, \quad x\in[0,\,1],
\end{equation} 
in fact, $f_n^P$ is obtained by replacing the histogram $h_{\w_{k_n}^{\mu_n}}$ with the frequency polygon $p_{\w_{k_n}^{\mu_n}}$ in \eqref{eq:Bestim}.
The smooth Barron-type density estimator $f_n^P$
can be given an interpretation in terms of the Bayes' rule
analogous to that of the Barron estimator $f_n^B$ presented by
\citet{BGvdM92} in Remark 5. Suppose that $X_1,\,\ldots,\,X_n$ are i.i.d. observations
from a distribution $F$ corresponding to a probability measure $P$ that is given a prior by assigning
a prior to the bin probabilities $(w_{1,\,k_n},\,\ldots,\,w_{k_n,\,k_n})=(P(A_{1,\,k_n}),\,\ldots,\,P(A_{k_n,\,k_n}))$
which is a Dirichlet distribution with parameters all equal to one, i.e.,
$\Pi_{k_n}=\mathrm{Dir}_{k_n}(1,\,\ldots,\,1)$.
Then, the posterior distribution of the cell probabilities, given the data, is still
Dirichlet with parameters $1+n\mu_n(A_{j,\,k_n})$, for $j=1,\,\ldots,\, k_n$.
Let $\w_{k_n}^{(n)}:=(w_{1,\,k_n}^{(n)},\,\ldots,\,w_{k_n,\,k_n}^{(n)})$, with $w_{j,\,k_n}^{(n)}$
the posterior expectation of the cell probability $w_{j,\,k_n}$, that is,
$w_{j,\,k_n}^{(n)}:=\mathds{E}[w_{j,\,k_n}\mid X^{(n)}]={[1+n\mu_n(A_{j,\,k_n})]}/({k_n+n})={(1+nw_{j,\,k_n}^{\mu_n})}/({k_n+n})$,
for $j=1,\,\ldots,\,k_n$, which may be interpreted as the relative frequency of the cell $A_{j,\,k_n}$
with one additional fictitious observation.
Then, the posterior expectation of a polygon constructed with
the bin probabilities $(w_{1,\,k_n},\,\ldots,\,w_{k_n,\,k_n})$ is
$\mathds{E}[p_{\w_{k_n}}(x)\mid X^{(n)}]
=n(k_n+n)^{-1}p_{\w_{k_n}^{\mu_n}}(x)+k_n(k_n+n)^{-1}=(1-a_n)p_{\w_{k_n}^{\mu_n}}(x)+a_n$, with $x\in[0,\,1]$.
Therefore, it is a convex combination of the polygonally smoothed empirical distribution function and
the prior guess which is the uniform distribution on $[0,\,1]$. Therefore,
$$f_n^P(x)=\mathds{E}[p_{\w_{k_n}}(x)\mid X^{(n)}],\quad x\in[0,\,1],$$
namely, the smooth Barron-type density estimator corresponds to the Bayes' estimator of a statistician who
assumes observations were generated from $F$ and takes a Dirichlet distribution with one \emph{a priori}
expected observation per cell as a prior for the cell probabilities.
In fact, in evaluating the expectation $\mathds{E}[p_{\w_{k_n}}(x)\mid X^{(n)}]$,
the posterior distribution of $P$ is computed assuming that $X_1,\,\ldots,\,X_n$ were i.i.d. observations from $F$.
A Bayesian statistician believing that the observations were generated from a density,
possibly a polygon, would, instead, first induce a prior on the space of polygon-shaped densities
from the prior distribution for
$P$ (or the mixing weights) and then compute the corresponding posterior.}

\emph{Barron's modification of the histogram estimator is motivated by the search for
consistency in stronger information divergence criteria than the $L^1$-distance,
which is needed for applications in information transfer and communication as illustrated in \citet{BGvdM92}.
The smooth Barron-type density estimator is in turn a modification of the Barron estimator to overcome discontinuities of the histogram.
The following result, which provides the order of the approximation error of any
density $f_0$, with Hölder regularity $\beta\in(0,\,2]$,
by the smooth Barron-type density estimator $f_n^P$ in the expected $\chi^2$-divergence, where $\chi^2(f_0\|f_n^P)\coloneqq
\int_0^1[(f_0-f_n^P)^2/f_n^P]\,\mathrm{d}\lambda$,
complements Theorem 4.1 of \citet{BBBV2002},
where only the case of a twice continuously
differentiable density is treated.
In what follows, we denote by $f^P_{n,\,\beta}$ the smooth Barron-type density estimator
corresponding to the choice $k_n=O(n^{1/(2\beta+1)})$.
\begin{proposition}\label{th2}
Let the density $f_0\in C^\beta([0,\,1])$, with $\beta\in(0,\,2]$, and $1/f_0\in L^1(\lambda)$.
For $\beta\in(1,\,2]$, suppose further that
$f_0'$ satisfies the boundary conditions $\mathrm{(BC1)}$ and $\mathrm{(BC2)}$.
Then,
$\mathds{E}_0^n[\chi^2(f_0\|f^P_{n})]=O(k_n^{-2\beta})+O(k_n/n)$.
The choice $k_n=O(n^{1/(2\beta+1)})$ gives
$\mathds{E}_0^n[\chi^2(f_0\|f^P_{n,\,\beta})]=O(\epsilon_{n,\,\beta}^2)$.
\end{proposition}
The next assertion provides a further aspect of the asymptotic behavior of the smooth Barron type density estimator.
Under some regularity conditions, at every point $x\in(0,\,1)$, the distribution of the re-scaled error made in estimating $f_0$
by the smooth Barron-type density estimator is asymptotically normal.
For $\alpha\in(0,\,1)$, let $z_{\alpha}$ be defined by $\textrm{P}(Z>z_\alpha)=\alpha$, with $Z\sim N(0,\,1)$.
\begin{proposition}\label{prop5}
Let the density $f_0$ be twice differentiable on $(0,\,1)$.
Suppose that $f_0,\,f_0',\,f_0''$ are bounded on $(0,\,1)$. For every $x\in(0,\,1)$,
choosing $k_n=O(n^{1/5})$,
\[\sqrt{\frac{n}{k_n}}[f_n^P(x)-f_0(x)]\rightarrow N\pt{\frac{1}{3!}f_0''(x),\,\frac{1}{2}f_0(x)}.\]
For any given $\alpha\in(0,\,1)$, the confidence interval
$[f_n^P(x)-z_{\alpha/2}n^{-2/5}\sqrt{f_0(x)/2},\,\,f_n^P(x)+z_{\alpha/2}n^{-2/5}\sqrt{f_0(x)/2}]$
is of asymptotic level less than or equal to $1-\alpha$.
\end{proposition}
The confidence interval cannot be immediately used in practice because the term $\sqrt{f_0(x)/2}$ depends on the
sampling density.}
\end{Ex}

\bigskip


So far, we have considered adaptation via sieve priors when a sequence of positive projection kernels is considered so that,
at each \vir{resolution} level, the Dirichlet process filtered through the kernel results in a density.
Another possibility is that of considering a \vir{convolution-type} kernel, like the Gaussian density,
with usual conversion from bin-width to bandwidth. Fully rate-adaptive density estimation
over locally Hölder density classes on the real line can be performed using finite
Dirichlet location mixtures of analytic exponential power densities as proposed
by \citet{KRvdV10}, where the sieve prior is obtained by first generating the
number of support points and next their locations and mixing weights according to a
Dirichlet distribution.
Mixture models with priors on the mixing distribution
admitting an infinite discrete representation, like the Dirichlet process or more general
stick-breaking priors, avoid choosing a truncation level for the number of mixing components.
Fully rate-adaptive density estimation over Sobolev or analytic regularity scales can be
performed using Dirichlet process mixtures of Gaussian densities as illustrated in the following example.

\begin{Ex} \emph{(}Gaussian mixtures\emph{)}.
\emph{The model is a location mixture
$f_P(\cdot)=f_{F,\,\sigma}(\cdot)=(F\ast \phi_\sigma)(\cdot)=\int_{-\infty}^\infty\sigma^{-1}\phi((\cdot-\theta)/\sigma)F(\mathrm{d}\theta)$,
where $\phi(\cdot)$ denotes the density of a standard Gaussian distribution, $\sigma$ the scale parameter and
$F$ the mixing distribution. Sampling densities $f_0$
herein considered are characterized via an integrated tail bound condition
on their Fourier transforms $\hat{f}_0(t)=\int_{-\infty}^\infty e^{itx}f_0(x)\,\mathrm{d}x$, $t\in\mathbb{R}$,
\begin{equation}\label{expdecr}
\int_{-\infty}^\infty (1+|t|^2)^\beta e^{2(\rho|t|)^r}|\hat{f}_0(t)|^2\,\mathrm{d}t\leq 2\pi L^2,
\end{equation}
for constants $0<\rho,\,L<\infty$, $\beta\in\mathbb{N}$ and $0\leq r<\infty$.
Densities with Fourier transforms satisfying condition (\ref{expdecr}) for $r>0$ 
constitute a larger collection than that of
analytic densities, including 
Gaussian, Cauchy, symmetric stable laws, Student's-$t$, distributions
with characteristic functions vanishing outside a compact set, as well as their mixtures and convolutions.
Densities with Fourier transforms satisfying condition (\ref{expdecr}) for $r=0$ are
called \emph{ordinary smooth}: they are differentiable up to the order $\beta$.
Examples of ordinary smooth distributions include gamma, double exponential and symmetric gamma
distributions. Given the model $f_{F,\,\sigma}$, a prior is induced on the space of Lebesgue densities by
putting priors on the mixing distribution $F$ and the scale parameter $\sigma$.
Let $\Pi$ denote the prior for $F$. The scale parameter is assumed to be distributed, independently
of $F$, according to a prior $G$ on $(0,\,\infty)$.
The sequence of observations $(X_i)_{i\geq1}$ is assumed to be exchangeable.
Observations from a kernel mixture prior can be described as
\[\begin{split}
X_i\mid (F,\,\sigma)\,&\overset{\textrm{iid}}{\sim}\,f_{F,\,\sigma},\quad i=1,\,\ldots,\,n,\\
(F,\,\sigma)\,&\overset{}{\sim}\,\Pi\times G.
\end{split}\]
The capability of convolution Gaussian kernel mixture priors to get optimal posterior contraction rates
depends on the approximation error of a density by Gaussian convolutions.
A well-known problem with the use of Gaussian convolutions is that the approximation error of a smooth density can only
be of the order $\sigma^2$, even if the density has greater smoothness, see for instance \citet{GvdV072b}.
The approximation can be improved using
higher-order kernels, but the resulting convolution is not guaranteed to be everywhere non-negative which,
in a frequentist approach, translates into a non-bona fide estimator, while it is not an issue in a Bayesian framework.
In fact, in the approach proposed in \citet{S2014}, which is reminiscent of that in \citet{KRvdV10},
the crux is the approximation of densities with Fourier transforms satisfying requirement
\eqref{expdecr} by convoluting the Gaussian kernel with an operator whose expression
is a series with suitably calibrated coefficients and density
derivatives that, in the supersmooth case, are further convoluted with the $\operatorname{sinc}$ kernel or, more generally, with a superkernel. This operation allows to reproduce the tail behavior of the Fourier transform of
$f_0$. Once this (not necessarily non-negative) function is suitably modified to be a density with the same tail behavior as $f_0$ and with the same
approximation properties in the sup-norm as well as in the Kullback-Leibler divergence,
the re-normalized restriction to a compact set of the corresponding continuous mixture is discretized
and a finite mixing distribution with a suitable number of support points such
that the corresponding Gaussian mixture is within \vir{small} Kullback-Leibler distance
from $f_0$ is found by matching a certain number of its moments with those of the previously
\emph{ad hoc} constructed mixing density. The key idea 
is that, under a set of regularity conditions on $f_0$ including \eqref{expdecr},
there exists a finite mixing distribution $F^*$ with $N_\sigma$ points in $[-a_\sigma,\, a_\sigma]$ such that
\begin{equation}\label{Holder}
\max\{\textrm{KL}(P_0;\,P^*_\sigma),\,\textrm{V}_2(P_0;\,P^*_\sigma)\}
\lesssim\sigma^{2\beta}
\mathds{1}_{\{0\}}(r)+\mathds{1}_{[1,\,2]}(r)e^{-c(1/\sigma)^{r}},
\end{equation}
where $P^*_\sigma$ is the probability measure corresponding to the density $f_{F^*,\,\sigma}$, the interval endpoint $a_\sigma = |\log \sigma|^{\rho_1}\mathds{1}_{\{0\}}(r)+\sigma^{-r/2}\mathds{1}_{[1,\,2]}(r)$ and $N_\sigma\lesssim (|\log\sigma|^{\rho_2}/\sigma)\mathds{1}_{\{0\}}(r)+(a_\sigma/\sigma)^2\mathds{1}_{[1,\,2]}(r)$.
Let $\epsilon_n^*=n^{-\beta/(2\beta+1)}\mathds{1}_{\{0\}}(r)+n^{-1/2}\mathds{1}_{[1,\,2]}(r)$.
\begin{theorem}\label{Th:DPMGaussian2}
Assume that $f_0$ satisfies conditions \eqref{expdecr} and \eqref{Holder}. Let the model be $f_{F,\,\sigma}=F\ast\phi_\sigma$.
Consider a prior distribution $\Pi\times G$ of the form $\mathrm{DP}(\alpha)\times G$, with
the base measure $\alpha$ of the Dirichlet process having a continuous and positive
density $\alpha'$ on $\mathbb{R}$ such that
$\alpha'(\theta)\propto e^{-b|\theta|^\delta}$ as $|\theta|\rightarrow\infty$,
for some constants $0<b<\infty$, $0<\delta\leq2$, and
$G=\mathrm{IG}(\nu,\,\lambda)$, with shape parameter $0<\nu<\infty$
and scale parameter $0<\lambda<\infty$. Then, for
$M>0$ large enough, $\mathbb{E}_0^n[
(\Pi\times G)((F,\,\sigma):\, \|f_{F,\,\sigma}-f_0\|_1> M\epsilon_n^*(\log n)^\kappa\mid
X^{(n)})]\rightarrow0$. Consequently, $\mathbb{E}^n_0[\|\hat{f}_n-f_0\|_1^2]=O((\epsilon_{n}^*)^2(\log n)^{2\kappa})$.
\end{theorem}
\citet{STG2013} have shown adaptation for multivariate (possibly anisotropic) locally H\"older regular densities
using a Dirichlet process mixture of normal densities, with a Gaussian base measure and
an inverse-Wishart prior on the covariance matrix, making use of the stick-breaking representation of the
Dirichlet process.
}
\end{Ex}


\section{Final remarks}\label{sec:3}
Bayesian adaptive estimation is increasingly better understood
in different statistical settings such as (conditional) density estimation,
regression and classification. The purpose of this article is to provide a survey of the main
approaches to Bayesian adaptation. As mentioned in the introduction,
the article deals with fully Bayes adaptation techniques only, but adaptation in an empirical Bayes
approach to inference has very recently begun to be deeply investigated. \citet{DRRS2014} provide general sufficient conditions in the spirit of
those proposed by \citet{GvdV072} to derive posterior contraction rates in models with general functional parameters which are then applied
to the specific setting of empirical Bayes adaptive density estimation and deconvolution using Dirichlet mixtures of Gaussian densities showing that
any data-driven choice of the prior base measure hyper-parameters lead to
minimax-optimal posterior contraction rates, up to logarithmic factors, provided the empirical Bayes hyper-parameter selection takes values
in a bounded set with high probability. We also refer the reader to the contribution of \citet{KSvdVvZ2012} where it is shown that
the maximum marginal likelihood selection of a hyper-parameter
related to the regularity level of the prior in the context of the inverse signal-in-white-noise model
leads to adaptive, rate-optimal procedures.

The article is focussed on point estimators, but in practice interest is in  quantifying uncertainty
by credibility regions, sets wherein the functional parameter takes values
with 
high posterior probability. The Bayesian approach to inference naturally produces
credibility regions, thus making Bayesian techniques appealing to practitioners.
However, the frequentist interpretation of these regions is still unclear and
needs to be carefully investigated 
in its many aspects. A credibility region is in fact a statement of probability about the functional parameter, given bounds
that depend on the observations. It is important to study frequentist validity of such bounds
in infinite-dimensional models because credibility regions
are not necessarily confidence sets as in the
finite-dimensional case, in the sense that, under the frequentist assumption that there exists
a true distribution generating the data, it is not automatically
guaranteed that they contain the true value of the parameter with
probability at least the level of the credibility region.
Early instances in the literature have pointed out that,
in different statistical settings, pairs of the true parameter values and priors may not match
in giving the right frequentist coverage of credibility regions.
As shown in \citet{KSvdVvZ2012}, Bayesian credibility regions typically have good frequentist coverage
when the prior is less regular than the true value of the parameter.
Since this is unknown, frequentist validity
of such bounds can be studied considering priors that
automatically adapt to unknown regularity by either taking an empirical Bayes approach that
employs a data-driven choice 
of the regularity level or a fully Bayes hierarchical approach with regularity hyper-parameter.
\citet{SvdVvZ2014} have results for empirical Bayes credibility regions
in the context of the inverse signal-in-white-noise model showing that there exist
values of the true parameter, forming a topologically small set in an appropriate sense, that are not covered by their
credibility regions. Empirical Bayes credible sets become adaptive confidence sets with
the right frequentist coverage if the \vir{unpleasant} values of the truth are removed.
Frequentist coverage of nonparametric credibility regions is an important topic
and is expected to be actively investigated in the near future.


\section*{Acknowledgments}
The author would like to thank the Editor and an anonymous referee for
influential reports whose constructive and meticulous suggestions helped
to substantially improve an earlier version of the manuscript.
Bocconi University is gratefully acknowledged for providing financial support.



\appendix

\section{Proof of Proposition \ref{prop3}}\label{App}

\begin{proof}[\bf Proof]
The result is known to hold true
for $\beta\in(0,\,1]$ from (13) in the proof of Theorem 4 in \citet{S2007}.
We first prove the assertion for $\beta\in(1,\,2]$.
We begin by showing that, under condition (BC1),
$\sup_{x\in A_{1,\,k}^-}|{f_0(x)-kw^0_{1,\,k}}|=
O(k^{-\beta})$.
For any fixed $x\in A_{1,\,k}^-$, by the Mean Value Theorem, for some $\xi\in(0,\,1/k)$, we have
$|{f_0(x)-kw^0_{1,\,k}}|
=|{f_0(x)-f_0(\xi)}|=|f'_0(\eta)||x-\xi|$, where $\eta$ is a point lying between $\xi$ and $x$. For $k$ large
enough so that $1/k$ is close to zero, by (BC1), $|{f_0(x)-kw^0_{1,\,k}}|<|f'_0(\eta)|/k=|a_0\eta^p+o(\eta^p)|/k<2|a_0|\eta^p/k$. For suitable $\delta\equiv\delta(\xi,\,x)\in(0,\,1)$, we
can write $\eta=\delta k^{-1}$. Then, $\sup_{x\in A_{1,\,k}^-}|{f_0(x)-kw^0_{1,\,k}}|<2|a_0|k^{-(p+1)}=O(k^{-\beta})$.
Analogously, using condition (BC2), we get that
$\sup_{x\in A_{k,\,k}^+}|{f_0(x)-kw^0_{k,\,k}}|=O(k^{-(q+1)})=O(k^{-\beta})$. For ease of notation,
we write $p_k^0$ as a short form for $p_{\w_k^0}$.
Next, we show that, for every $j=1,\,\ldots,\,k-1$,
\begin{equation}\label{eq10}
\sup_{x\in A_{j,\,k}^+\cup
A^-_{j+1,\,k}}|{f_0(x)-p_k^0(x)}|=O(k^{-\beta}).
\end{equation}
Write $w^0_{j,\,k}=F_0(j/k)-F_0((j-1)/k)$, where $F_0$ is the cumulative distribution function 
of the density $f_0$. 
A second-order Taylor expansion of $F_0((j-1)/k)$ and $F_0(j/k)$ near
$c_{j,\,k}$, with the remainder term in the Lagrange form, yields that
for points $\zeta\in((j-1)/k,\,c_{j,\,k})$ and
$\zeta'\in(c_{j,\,k},\,j/k)$,
\begin{equation}\label{eq8}
w^0_{j,\,k}=F_0(c_{j,\,k})+\frac{1}{2k}f_0(c_{j,\,k})+\frac{1}{8k^2}f_0'(\zeta')
-\pq{F_0(c_{j,\,k})-\frac{1}{2k}f_0(c_{j,\,k})+\frac{1}{8k^2}f_0'(\zeta)}=\frac{1}{k}f_0(c_{j,\,k})+\frac{1}{8k^2}[f_0'(\zeta')
-f_0'(\zeta)].
\end{equation}
By the same argument, for points $\theta\in (j/k,\,c_{j+1,\,k})$ and
$\theta'\in(c_{j+1,\,k},\,(j+1)/k)$,
\begin{eqnarray}\label{eq9}
w^0_{j+1,\,k}=\frac{1}{k}f_0(c_{j+1,\,k})+\frac{1}{8k^2}[f_0'(\theta')
-f_0'(\theta)].
\end{eqnarray}
For later use, note that $[(\zeta'-\zeta)\vee(\theta'-\theta)]<k^{-1}$. For any $x\in A_{j,\,k}^+\cup A_{j+1,\,k}^-$, the value
of the density $p_k^0(x)$ can be neatly written in the form
$p_k^0(x)=q_{j+1,\,k}(x)kw^0_{j,\,k}+[1-q_{j+1,\,k}(x)]kw_{j+1,\,k}^0$,
where we set $q_{j+1,\,k}(x):= k(c_{j+1,\,k}-x)$ and, consequently,
$1-q_{j+1,\,k}(x)=k(x-c_{j,\,k})$. Using the expressions in
(\ref{eq8}) and (\ref{eq9}), together with the assumption that $f_0'$
has Hölder regularity $(\beta-1)$, we get that, for points $\vartheta\in(c_{j,\,k},\,x)$
and $\vartheta'\in(x,\,c_{j+1,\,k})$,
\begin{eqnarray*}
\begin{split}
|f_0(x)-p_k^0(x)|&=\abs{q_{j+1,\,k}(x)[f_0(x)-k w_{j,\,k}^0]+[1-q_{j+1,\,k}(x)][f_0(x)-k w_{j+1,\,k}^0]}
\\&\leq\abs{q_{j+1,\,k}(x)f_0'(\vartheta)\frac{1-q_{j+1,\,k}(x)}{k}-[1-q_{j+1,\,k}(x)]
f_0'(\vartheta')\frac{q_{j+1,\,k}(x)}{k}}\\
&\hspace*{6.7cm} + \frac{q_{j+1,\,k}(x)}{8k}L_0|\zeta'-\zeta|^{\beta-1}
+\frac{1-q_{j+1,\,k}(x)}{8k}L_0|\theta'-\theta|^{\beta-1}\\
&<\frac{q_{j+1,\,k}(x)[1-q_{j+1,\,k}(x)]}{k}|f'_0(\vartheta)-f'_0(\vartheta')|+\frac{L_0}{8}k^{-\beta}\\
&< q_{j+1,\,k}(x)[1-q_{j+1,\,k}(x)]L_0k^{-\beta}+\frac{L_0}{8}k^{-\beta}\\&\leq\frac{3L_0}{8}
k^{-\beta}
\end{split}
\end{eqnarray*}
and \eqref{eq10} follows. Lastly, we consider the case where $\beta>2$. By the same reasoning as before,
$\sup_{x\in A_{1,\,k}^-}|{f_0(x)-kw^0_{1,\,k}}|=
O(k^{-\beta})$ and $\sup_{x\in A_{k,\,k}^+}|{f_0(x)-kw^0_{k,\,k}}|=O(k^{-\beta})$.
Since $f'_0$ is differentiable on $[0,\,1]$ and $f''_0$ is bounded, $f'_0$ is Lipschitz continuous
with constant $L_0=\max_{x\in[0,\,1]}|f'_0(x)|$. Hence, by the same arguments as above,
for every $j=1,\,\ldots,\,k-1$, we have $\sup_{x\in A_{j,\,k}^+\cup
A^-_{j+1,\,k}}|{f_0(x)-p_k^0(x)}|=O(k^{-2})$. Thus, $\|f_0-p_k^0\|_\infty=O(k^{-(\beta\wedge2)})=O(k^{-2})$
and the proof is complete.
\end{proof}


\section{Proof of Proposition \ref{prop4}}
\begin{proof}[\bf Proof]
We first sketch the underlying reasoning.
Let $\tilde\epsilon_n$ and $\zeta_n$ be positive sequences
such that $(\tilde\epsilon_n\vee\zeta_n)\rightarrow0$ and $n(\tilde\epsilon_n^2\wedge\zeta_n^2)\rightarrow\infty$.
Both sequences will be specified below. Let $P_0$ denote the probability law having Radon-Nikodym derivative
$f_0$ with respect to Lebesgue measure $\lambda$, i.e., $f_0=\mathrm{d}P_0/\mathrm{d}\lambda$. 
It is known that if, for some constant $C>0$, the prior mass
$\Pi(B_{\textrm{KL}}(P_0;\,\tilde\epsilon_n^2))\geq e^{-Cn\tilde\epsilon_n^2}$, then, for any measurable set $A$,
\begin{equation}\label{CP}
\forall\,\eta>0,\quad P_0^{n}(\Pi(A\mid X^{(n)})>\eta)\lesssim e^{(1+2C)n\tilde\epsilon_n^2}\Pi(A)+o(1).
\end{equation}
If $e^{(1+2C)n\tilde\epsilon_n^2}\Pi(A)=o(1)$, 
then 
the posterior probability of $A$ is negligible.
Define the set $A_{\zeta_n}(P_0):=\{P:\,\|f_P-f_0\|_1\leq\zeta_n\}$. In view of \eqref{CP},
in order to show that $\zeta_n$ is a lower bound on the posterior $L^1$-contraction rate at $f_0$,
it is enough to show that $\Pi(A_{\zeta_n}(P_0))\lesssim e^{-c_1 n^\kappa}$ for some constant $c_1>0$ and an exponent
$\kappa>\ell>0$, where $\ell$ is such that $n^\ell=n\tilde\epsilon_n^2$. It then follows that
$\Pi(A_{\zeta_n}(P_0)\mid X^{(n)})\rightarrow0$ in $P_0^n$-probability.
In the specific setting of Proposition \ref{prop4}, as it will be shown below, we have that
for every sufficiently large $k$,
\begin{equation}\label{LB}
\inf_{\w_k \in \Delta^{k-1}}\|p_{\w_k}-f_0\|_1\gtrsim k^{-3/2}.
\end{equation}
Let $\zeta_n=\epsilon_{n,\,3}=n^{-3/7}$. In view of \eqref{LB}, in order for a polygon
$p_{\w_k}$ to be in $A_{\zeta_n}(P_0)$, we need that
$k\gtrsim \zeta_n^{-2/3}$. So, for a suitable constant $c_1>0$, we have
$\Pi(A_{\zeta_n}(P_0))\leq \textrm{P}(k\gtrsim \zeta_n^{-2/3})\lesssim e^{-c_1n^{2/7}}$.
Since $\tilde{\epsilon}_n=(n/\log n)^{-2/5}$, we have $\kappa=2/7>1/5=\ell$, which implies that
$\zeta_n$ is a lower bound on the posterior contraction rate at $f_0$.

We now prove \eqref{LB}. We partly follow the lines of the proof
of Lemma 4.2 in \citet{McVinishRousseauMengersen2005}.
For $k$ large enough, there exists an index $j\in\{1,\,\ldots,\,k\}$ such that $A_{j,\,k}\subset I$.
We consider the case where $f_0''<0$ on $I$. The same arguments apply if
$f_0''>0$. Let $p_k^*$ be such that $\|p_k^*-f_0\|_1=\inf_{\w_k \in \Delta^{k-1}}\|p_{\w_k}-f_0\|_1$. Then, $\|p_k^*-f_0\|_1\geq\int_{A_{j,\,k}}|p_k^*(x)-f_0(x)|\,\mathrm{d}x>(\int_{A_{j,\,k}}|(a^*+b^*x)-f_0(x)|\,\mathrm{d}x)^{1/2}$, where
$a^*,\,b^*$ minimize the last expression. As $f_0$ is strictly concave on $I$ (hence on $A_{j,\,k}$), the line
$a^*+b^*x$ intersect $f_0$ at two points $x_1,\,x_2\in A_{j,\,k}$. Let $\bar x=(x_1+x_2)/2$. Then, $\|p_k^*-f_0\|_1^2$ is bounded below by
the area of the triangle formed by the points
of coordinates $(x_1,\,f_0(x_1))$, $(\bar x,\, f_0(\bar x))$ and $(x_2,\,f_0(x_2))$ which is equal to
$$\frac{1}{2}(x_2-x_1)\left[f_0(\bar x)-\frac{f_0(x_1)+f_0(x_2)}{2}\right]=
\frac{1}{16}(x_2-x_1)^3[-f''_0(x_2)+O(x_2-x_1)]=\frac{\delta^3}{16}[-f''_0(x_2)+O(x_2-x_1)]k^{-3}
$$
because $(x_2-x_1)=\delta k^{-1}$ for $\delta\equiv \delta(x_1,\,x_2)\in(0,\,1)$. It follows that
$\|p_k^*-f_0\|_1\gtrsim k^{-3/2}$ and the proof is complete.
\end{proof}


\section{Proof of Proposition \ref{th2}}

\begin{proof}[\bf Proof]
It is known from Theorem 4.1 of \citet{BBBV2002} that
$\mathds{E}_0^n[\chi^2(f_0\|f_n^P)]\leq \chi^2(f_0\|p^0_{k_n})+
[1+\chi^2(f_0\|p^0_{k_n})]k_n/(n+1)$. By Proposition \ref{prop3}
and the assumption that $1/f_0\in L^1(\lambda)$, we have
$\chi^2(f_0\|p^0_{k_n})=O(k_n^{-2\beta})$. It follows that
$\mathds{E}_0^n[\chi^2(f_0\|f_n^P)]=O(k_n^{-2\beta})+O(k_n/n)$.
By choosing $k_n=O(n^{1/(2\beta+1)})$, we have
$\mathds{E}_0^n[\chi^2(f_0\|f_{n,\,\beta}^P)]=O(\epsilon_{n,\,\beta}^2)$.
\end{proof}

\section{Proof of Proposition \ref{prop5}}
\begin{proof}[\bf Proof]
Let $x\in(0,\,1)$ be fixed. Write
\[\sqrt{\frac{n}{k_n}}[f_n^P(x)-f_0(x)]=(1-a_n)\sqrt{\frac{n}{k_n}}
[p_{\w_{k_n}^{\mu_n}}(x)-f_0(x)]+a_n\sqrt{\frac{n}{k_n}}[\mathds{1}_{[0,\,1]}(x)-f_0(x)].\]
Since $f_0$ is bounded on $(0,\,1)$ and $a_n(n/k_n)^{1/2}=(1+n/k_n)^{-1}(n/k_n)^{1/2}\rightarrow0$,
the second term on the right-hand side of the above identity $a_n({n}/{k_n})^{1/2}[\mathds{1}_{[0,\,1]}(x)-f_0(x)]\rightarrow0$ as $n\rightarrow\infty$.
We study the first term. Since $a_n\rightarrow0$, the factor $1-a_n$ can be neglected.
Consistently with the notation introduced in the proof of Proposition \ref{prop3},
we write $p_{k_n}^0$ as a short form for $p_{\w_{k_n}^0}$ and we have
\begin{equation}\label{gaus}
\sqrt{\frac{n}{k_n}}
[p_{\w_{k_n}^{\mu_n}}(x)-f_0(x)]=\sqrt{\frac{n}{k_n}}[p_{\w_{k_n}^{\mu_n}}(x)-p_{k_n}^0(x)]+\sqrt{\frac{n}{k_n}}
[p_{k_n}^0(x)-f_0(x)].
\end{equation}
Hereafter, we study the asymptotic behavior of the two terms on the right-hand side of \eqref{gaus}.
\begin{description}
\item[$(i)$] \emph{Study of the term $\sqrt{\dfrac{n}{k_n}}[p_{\w_{k_n}^{\mu_n}}(x)-p_{k_n}^0(x)]$.}\\
We partly follow the lines of \citet{G2001}.
Let $F_n(x)=n^{-1}\sum_{i=1}^n\mathds{1}_{(-\infty,\,x]}(X_i)$, $x\in\mathbb{R}$, be the empirical distribution function associated with the sample of i.i.d. observations $X_1,\,\ldots,\,X_n$ from $F_0$, where $F_0$ denotes the cumulative distribution function of the probability law $P_0$ having
Radon-Nikodym derivative $f_0$ with respect to Lebesgue measure $\lambda$, i.e., $f_0=\mathrm{d}P_0/\mathrm{d}\lambda$.
By the result of \citet{KMT}, 
$\sqrt{n}[F_n(x)-F_0(x)]$ is uniformly approximated by a Brownian bridge
$B_n(F_0(x))$ almost surely with an error of the order $n^{-1/2}\log n$. By a well-known result, we can write
$B_n(F_0(x))=W_n(F_0(x))-F_0(x)W_n(1)$, where $W_n(t)$ is a Wiener process. Therefore,
\begin{equation}\label{eq:decomp}
\sqrt{\frac{n}{k_n}}[p_{\w_{k_n}^{\mu_n}}(x)-p_{k_n}^0(x)]= T_n(x) -k_n^{-1/2}p_{k_n}^0(x)W_n(1)+O(k_n^{-1/2}n^{-1/2}\log n),
\end{equation}
where, using the notation $q_{j+1,\,k_n}(x):=k_n(c_{j+1,\,k_n}-x)$ introduced
in the proof of Proposition \ref{prop3},
\[
\begin{split}
T_n(x)&:= k_n^{-1/2}
[W_n(F_0(1/k_n))-W_n(F_0(0))]k_n\mathds{1}_{A_{1,\,k_n}^-}(x)\\
&\qquad\qquad\qquad\quad +k_n^{-1/2}\sum_{j=1}^{k_n-1}
q_{j+1,\,k_n}(x)[W_n(F_0(j/k_n))-W_n(F_0((j-1)/k_n))]k_n\mathds{1}_{A_{j,\,k_n}^+\cup
A^-_{j+1,\,k_n}}(x)\\
& \qquad\qquad\qquad\quad +k_n^{-1/2}\sum_{j=1}^{k_n-1}[1-q_{j+1,\,k_n}(x)][W_n(F_0((j+1)/k_n))-W_n(F_0(j/k_n))]
k_n\mathds{1}_{A_{j,\,k_n}^+\cup
A^-_{j+1,\,k_n}}(x)\\
&\qquad\qquad\qquad\quad +k_n^{-1/2}
[W_n(F_0(1))-W_n(F_0(1-1/k_n))]k_n\mathds{1}_{A_{k_n,\,k_n}^+}(x)\\
&=: T_n^{(1)}(x) + T_n^{(2)}(x) + T_n^{(3)}(x) + T_n^{(4)}(x).
\end{split}
\]
We analyze the terms $T_n(x)$ and $k_n^{-1/2}p_{k_n}^0(x)W_n(1)$ appearing in \eqref{eq:decomp}.
We begin by showing that $$T_n(x)\overset{d}{\rightarrow} N\pt{0,\,\frac{1}{2}f_0(x)}.$$
Since $x$ is fixed, for every $n$ large enough, $x\in \bigcup_{j=1}^{k_n-1}(A_{j,\,k_n}^+\cup
A^-_{j+1,\,k_n})$. Thus, $T_n^{(1)}(x)=T_n^{(4)}(x)=0$.
We study $T_n^{(2)}(x)+T_n^{(3)}(x)$. 
By definition of a Wiener process, $W_n(0)=0$ (hence, $W_n(F_0(0))=W_n(0)=0$), the increments $[W_n(F_0(1/k_n))-W_n(F_0(0))],\,\ldots,\,
[W_n(F_0(1))-W_n(F_0(1-1/k_n))]$ are independent random variables and
$[W_n(F_0(j/k_n))-W_n(F_0((j-1)/k_n))]\sim N(0,\,w_{j,\,k_n}^0)$, $j=1,\,\ldots,\,k_n$. Furthermore,
$f_0'$ and $f_0''$ are bounded on $(0,\,1)$
(hence, $f_0'(x)<\infty$ and $f_0''(x)<\infty$). Thus,
\[T_n^{(2)}(x)+T_n^{(3)}(x)\sim N\pt{0,\,\sum_{j=1}^{k_n-1}\pg{
q^2_{j+1,\,k_n}(x)w_{j,\,k_n}^0+[1-q_{j+1,\,k_n}(x)]^2w_{j+1,\,k_n}^0}k_n\mathds{1}_{A_{j,\,k_n}^+\cup
A^-_{j+1,\,k_n}}(x)}\overset{d}{\rightarrow} N\pt{0,\,\frac{1}{2}f_0(x)}.\]
We now prove that
$$k_n^{-1/2}p_{k_n}^0(x)W_n(1)=O_p(k_n^{-1/2}).$$
Since $W_n(1)\sim N(0,\,1)$ and, as subsequently shown, $p_{k_n}^0(x)\rightarrow f_0(x)$ at each point $x\in(0,\,1)$, the term $k_n^{-1/2}p_{k_n}^0(x)W_n(1)=O_p(k_n^{-1/2})$.
To prove that $p_{k_n}^0(x)\rightarrow f_0(x)$ on $(0,\,1)$, we write $p_{k_n}^0(x)-f_0(x)=[p_{k_n}^0(x)-h_{k_n}^0(x)]+[h_{k_n}^0(x)-f_0(x)]$,
where $h_{k_n}^0$ is a short form for $h_{\w_{k_n}^0}$. The assertion then follows from the convergence
\begin{equation}\label{eq11}
p_{k_n}^0(x)-h_{k_n}^0(x)\rightarrow0,
\end{equation}
together with the well-known companion
result $h_{k_n}^0(x)\rightarrow f_0(x)$. To see \eqref{eq11}, write
\[\begin{split}
0\leq |h_{k_n}^0(x)-p_{k_n}^0(x)|
&=k_n\sum_{j=1}^{k_n-1}
\{[1-q_{j+1,\,k_n}(x)]\mathds{1}_{A_{j,\,k_n}^+}(x)+q_{j+1,\,k_n}(x)\mathds{1}_{A^-_{j+1,\,k_n}}(x)\}|w_{j,\,k_n}^0-w_{j+1,\,k_n}^0|\\
&\leq k_n\sum_{j=1}^{k_n-1}\{[1-q_{j+1,\,k_n}(x)]\mathds{1}_{A_{j,\,k_n}^+}(x)+q_{j+1,\,k_n}(x)\mathds{1}_{A^-_{j+1,\,k_n}}(x)\}\int_0^{1/k_n}|f_0(j/k_n+t)-f_0(j/k_n-t)|\,\mathrm{d}t.
\end{split}\]
The density $f_0$ is continuous and has bounded derivative on $(0,\,1)$, hence, it is uniformly continuous, i.e.,
for any given $\epsilon>0$ there exists $\delta_\epsilon>0$ such that, whenever
$|x-y|<\delta_\epsilon$, we have $|f_0(x)-f_0(y)|<\epsilon$. For $k_n>k_\epsilon:=2/\delta_\epsilon$, we have $|(j/k_n+t)-(j/k_n-t)|=2t\leq2/k_n<\delta_\epsilon$ so that
$|f_0(j/k_n+t)-f_0(j/k_n-t)|<\epsilon$. Consequently,
$|h_{k_n}^0(x)-p_{k_n}^0(x)|<\epsilon$. Thus, for any given
$\epsilon>0$, there exists $k_\epsilon>0$ such that
$|h_{k_n}^0(x)-p_{k_n}^0(x)|<\epsilon$ for all $k_n>k_\epsilon$.
\item[$(ii)$] \emph{Study of the term $\sqrt{\dfrac{n}{k_n}}
[p_{k_n}^0(x)-f_0(x)]$.}\\ As before, for every $n$ large enough, $x\in\bigcup_{j=1}^{k_n-1}(A_{j,\,k_n}^+\cup
A^-_{j+1,\,k_n})$. Straightforward computations lead to $[p_{k_n}^0(x)-f_0(x)]\sim k_n^{-2}f_0''(x)/3!$,
where, by writing $a_n\sim b_n$ ($n\rightarrow\infty$), we mean that $b_n\neq 0$ and $\lim_{n\rightarrow\infty}(a_n/b_n)=1$.
For $k_n=n^{1/5}$, we have $(n/k_n)^{1/2}=k_n^2$ and
\[
\sqrt{\frac{n}{k_n}}
[p_{k_n}^0(x)-f_0(x)]\sim \frac{1}{3!}f_0''(x).
\]
\end{description}
Combining partial results in $(i)$ and $(ii)$, we have
\[
\sqrt{\frac{n}{k_n}}
[p_{\w_{k_n}^{\mu_n}}(x)-f_0(x)]\rightarrow N\pt{\frac{1}{3!}f_0''(x),\,\frac{1}{2}f_0(x)}
\]
and the proof is complete.
\end{proof}









\nocite{*}
\bibliographystyle{plain}



\end{document}